\documentclass[a4paper,twoside]{article}
\newcommand{\affil}[1]{$^{\rm #1}$}
\newcommand{\ltsima} {$\; \buildrel < \over \sim \;$}
\newcommand{\gtsima} {$\; \buildrel > \over \sim \;$}
\newcommand{\lta} {\lower.5ex\hbox{\ltsima}}
\newcommand{\gta} {\lower.5ex\hbox{\gtsima}}
\newcommand{\HI}{H{\,{\sc i}}}
\newcommand{\cena}{Centaurus~A}

\input{psfig}
\usepackage[authoryear]{natbib}
\bibpunct{(}{)}{;}{a}{}{,}
\date{} 
\newcommand{\kms}{\mbox{km\,s$^{-1}$}}
\title{\large\bf\flushleft The many faces of the gas in Centaurus A  (NGC~5128)}
\author{\parbox{\textwidth}{\flushleft
\vspace{-0.5cm}
{\it Raffaella Morganti\affil{1,2,*}}\\
\vspace{0.4cm}
{\small \affil{1}\,Netherlands Institute for Radio Astronomy, Postbus 2,
7990 AA, Dwingeloo, The Netherlands}\\
{\small \affil{2}\,Kapteyn Astronomical Institute, University of Groningen, P.O. Box 800,
9700 AV Groningen, The Netherlands}\\
{\small \affil{*}\,Email: morganti@astron.nl}}}
\begin{document}
\twocolumn[
\begin{changemargin}{.8cm}{.5cm}
\begin{minipage}{.9\textwidth}
\vspace{-1cm}
\maketitle
%
%
\small{\bf Abstract:}

\cena\ (NGC~5128) is a fantastic object, ideal for investigating the characteristics and the r\^ole of the gas in an early-type galaxy in the presence of a radio-loud active nucleus.  The different phases of the gas - hot (X-ray), warm (ionised) and cold (\HI\ and molecular) - are all detected in this object  and can be studied, due to its proximity,  at very  high spatial resolution. 
This richness makes \cena\  truly  unique.
Spatially, these gas structures span from the pc to the tens of kpc scale. Thus, they allow us to trace very different phenomena, from  the  formation and evolution of the host galaxy, to  the interplay between  nuclear activity and ISM and  the feeding mechanism of the central black hole. 
A lot of work has been done to study and understand the characteristics of the gas in this complex object  and here I summarise  what has been achieved so far.


\medskip
\medskip
\end{minipage}
\end{changemargin}
]
\small

\section{Gas in Centaurus~A: why interesting?}

The last years have seen  a change of paradigm in our view of early-type galaxies. Recent detailed observations of these objects in different wavebands, have confirmed and emphasised the picture - already suggested in earlier studies (see e.g.  Kormendy \& Bender 1996; Faber et al. 1997 and ref. therein) -  that the structure, dynamics, stellar populations and nuclear activity of early-type galaxies is complex and therefore early-type galaxies do not  constitute a uniform class. 
In addition to this, it has become clear that gas is an important ingredient even in these objects, an element  that cannot be neglected (see e.g. Bender, Burstein \& Faber 1992, Cappellari et al. 2007 and refs. therein). Recent observations, deeper than available before,  have shown that they do have a rich, complex and multiphase interstellar medium which contains hot (X-ray), warm (ionised) and cold (\HI\ and molecular) components.  We know that early-type galaxies can contain large quantities - up to at least several $10^7$ M$_\odot$ - of hot gas (see Mathews \& Brigenti 2003 for a review). However, for the remaining phases of the gas and in particular cold gas, early-type galaxies used to be perceived as gas poor. Although, indeed, they typically have less cold gas than spiral galaxies, it is now clear that cold gas is detected in many early-type galaxies, provided  that deep observations are available (see e.g. Sarzi et al. 2006, Combes et al. 2007, Morganti et al. 2006). The  variety of characteristics of their gaseous structures  can be considered as a signatures of the continuing assembly of these objects. The relevance of gas in the formation and evolution of early-type galaxies is
also suggested by theoretical work that indicates that dissipative mergers and accretion events are needed to explain the dynamical structure of, in particular, the more disky early-type galaxies  (see e.g. Naab \& Burkert 2003, Bournaud, Jog, \& Combes 2005,  Hopkins et al. 2009). 

Inside this revised picture of early-type galaxies, \cena\ represents a nice example of the complexity described above. \cena\  (NGC~5128)$\footnote{To be correct, I should be using NGC~5128 when discussing the properties of the optical galaxy, however for convenience, but actually mostly for habit, I will be using  \cena\ - that is in fact the name of the radio source - throughout this paper.}$  is  an early-type galaxy (see Harris et al. 2010) with a  minor axis dust-lane and a  major axis stellar component, therefore showing a strong misalignment between stars and gas,  relatively common in early-type galaxies. \cena\  has the exciting property of showing the presence of different phases of the gas with structures spread across a wide range of spatial  scales, going from the sub-pc scale up to tens of kpc. 
Thanks to its proximity, \cena\ offers the possibility of exploring the  structure and kinematics of the gas  with very  high spatial resolution. This, combined with the fact that \cena\ is an early-type galaxy hosting a radio-loud AGN, makes this object truly  unique. 

A lot of work has been done to study and understand the characteristics of the gas in \cena. 
The  gas has been used to trace different phenomena and help in understanding  the characteristics of this galaxy. 
In particular - and more related to the case of an active nucleus like \cena\ - the study of the gas allows us to:
\begin{itemize}
\item   probe  the radial mass distribution and the  shape of early-type galaxies. This provides also information on 
conditions (and time-scales) during their formation;  
\item trace signatures of  mergers/interactions or recent accretions, again giving information about the formation history of the galaxy;  
\item understand the nuclear structure, in particular  through the study of gas kinematics.  This  provides information on the fuelling mechanism, a still open question especially  for relatively low-power radio galaxies like \cena;  
\item trace the presence of outflows and the effects of AGN-related feedback   
\end {itemize}
 
 \begin{figure}
\begin{center}
\centerline{\psfig{figure=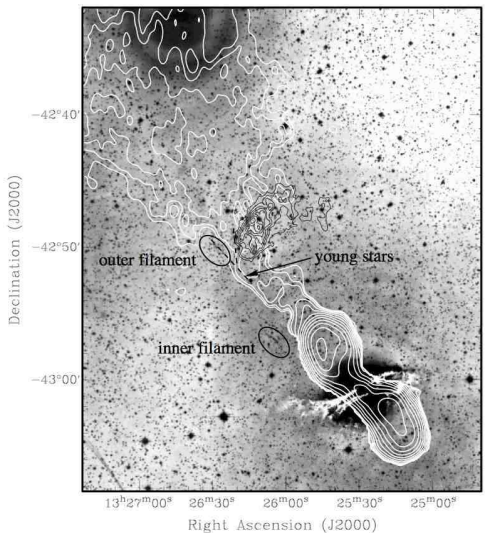,width=7cm,angle=0}}
\caption{Overlay showing the various components described in the text. The optical image (kindly provided by D.Malin) shows 
the well-known dust lane of \cena\ and the faint  diffuse  optical emission that extends to large distances from the centre.  The white contours denote the 
radio continuum emission (after Morganti et al. 1999a) showing the bright inner radio lobes and the large-scale jet that connects these lobes to  the so-called Northern Middle Lobe  in the top-left of the figure. The black contours denote the \HI\  cloud discussed in this paper (see  also Fig. 3). The locations of the inner and outer filaments of highly ionised gas are indicated, as well as the location of young stars.}\label{figfilaments}
\end{center}
\end{figure}

It is worth noting that \cena\ has clear evidence (dust-lane, shells etc.) that the gas has been  brought in through the accretion of a small gas-rich galaxy. This is  not always clear-cut in other early-type galaxies. 
In fact, although mergers can provide a good description of many characteristics of early-type galaxies, it is nevertheless important to keep in mind that gas can also be acquired in other ways. Particularly interesting is the cold accretion, the slow but long-lasting infall of primordial gas (see e.g., Keres et al.\ 2005\nocite{2005MNRAS.363....2K}, Dekel et al. 2009).
Thus, in the case of \cena\ we can really study the effects of merger in shaping the characteristics of the galaxy, including the presence of a radio-loud AGN.

This paper is an attempt to give an overview of the many faces of the gas in \cena\  although, given the extensive (huge!) literature available on this subject, it will be impossible to be complete.  The rest of the paper describes the gaseous structures in \cena\ starting from the large scale and then zooming-in  toward the inner, nuclear regions. The paper starts describing the  gas observed up to $\sim 20$ kpc  from the centre. After that, it will move to the gas observed along the dust lane (up to $\sim 6$ kpc from the centre). Finally,  it will explore the sub-kpc and the nuclear ($< 30$ pc) regions to investigate possible effects  of the radio jet and of the black hole (BH). Furthermore, the paper looks at the possible effects of the interaction between the large scale radio emission and the  gas. For a more general review of \cena\ - NGC~5128 see Israel (1998).

\begin{figure}
\begin{center}
\centerline{\psfig{figure=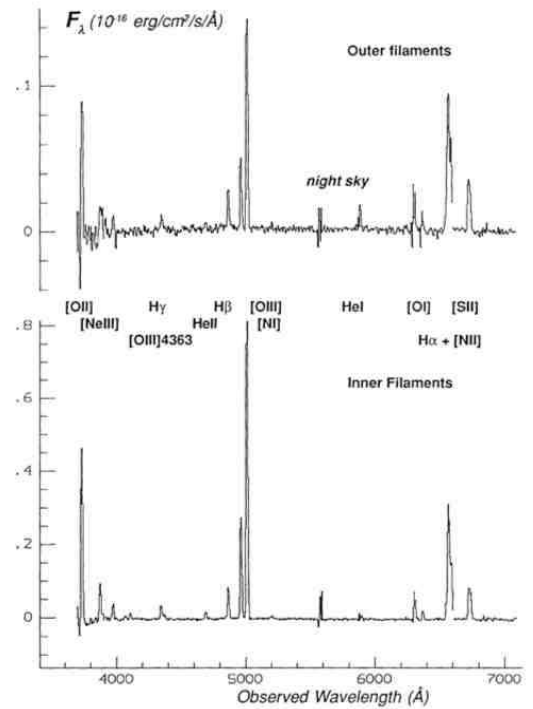,width=7cm,angle=0}}
\caption{An example of the inner and outer filament spectra taken from Morganti et al. (1991). The principal emission lines are labelled.}\label{figspectrum}
\end{center}
\end{figure}

\section{Gas in \cena: structures at large scale }

At large distances from the centre, the ISM of \cena\ shows a complex and intriguing structure. 
Most of the action happens in the NE side of the galaxy,  the same side where the most complex part of the radio emission is also observed (the so-called north-middle lobe and the outer lobe). Whether there is a connection between these two facts, e.g. due to the presence of a particularly dense external medium at this location (see the results on the hot gas at the end of this session), is not yet clear. The location of the various gas structures is illustrated in Fig. \ref{figfilaments}.

Highly ionised filaments of gas have been observed in this region, extending  up to at least 20 kpc from the nucleus to the northern side of the galaxy. The extended filaments were discovered by Blanco et al. (1975): the {\sl inner filament} located at a projected distance from the centre of 7 kpc and  extended $\sim 2$ kpc and the {\sl outer filament} located at a projected distance of 16~kpc. In both structures (but in particular in the inner filament) the gas shows very high ionisation. An example of optical spectrum  is shown in Fig. \ref{figspectrum}. 
More closer to the centre,  Dufour \& van den Bergh (1978)  pointed out the presence of an {\sl inner optical jet}  made of compact knots and diffuse filaments but this structure was not confirmed by narrow band images (Morganti et al. 1991).  Interestingly, no similar filaments have been found in the southern part of the galaxy.  

The emission line spectra  of the filaments are very similar in character to those of the spatially extended nebulosities in more distant radio galaxies and radio quasars. From the analysis of these lines, Morganti et al. (1991) concluded that the filaments are photoionised by the radiation field of a nuclear continuum source which is hidden from our direct view either by obscuration or by intrinsic anisotropy (or both). This hypothesis is supported by the ionisation gradient (i.e. higher ionisation on the side closer to the nucleus) observed in the inner filament (Morganti et al. 1992). The energy required to photoionise the gas would be consistent with \cena\ having a beam power similar to that of BL~Lac.  The similarity between  \cena\ and BL Lac is also confirmed by the near-IR study and the study of the spectral energy distribution by  Marconi et al. (2000).

However, in addition to the high excitation, the gas in the filaments shows velocity variations of 100 -- 200 \kms\ happening over regions of only a few hundred pc.  This complex kinematics cannot be accounted for by radiation pressure.  Particle beam acceleration seems to be necessary, thus suggesting  the possibility (explored by Sutherland et al. 1993) that  the ionisation of the filaments is actually connected to shocks produced by the interaction of the radio jet with the ISM.  However, at least  the inner filament does not appear to experience (at least by looking at the radio image) a direct interaction with the radio jet  that would produce fast enough shocks and, consequently,  the high ionisation observed (see Morganti et al. 1999). 
Thus, it is more likely that a combination of  both mechanisms provides the observed characteristics. In this respect, the region where the outer filaments are located is  particularly interesting. The radio emission (in the form of a collimated structure, see Morganti et al. 1999) appears, in projection, to pass next to the outer filaments  of highly ionised gas and next to a large cloud/tail of neutral gas (Schiminovich et al.\ 1994); see Fig. \ref{figfilaments} for an overview of the location of these structures.
The location of the \HI\ next to the filaments of ionised gas suggests that a narrow radiation beam could be the origin of the ionisation. However, higher velocity resolution \HI\ observations (Oosterloo \& Morganti 2005) have revealed that, apart from the smooth velocity gradient corresponding to the overall rotation of the \HI\ cloud/tail around Centaurus A, \HI\ with anomalous velocities of about 100 \kms\ is present at the southern tip of this cloud. This has been interpreted as evidence for an ongoing interaction between the radio plasma and the \HI\ cloud.  

\begin{figure}
\begin{center}
\centerline{\psfig{figure=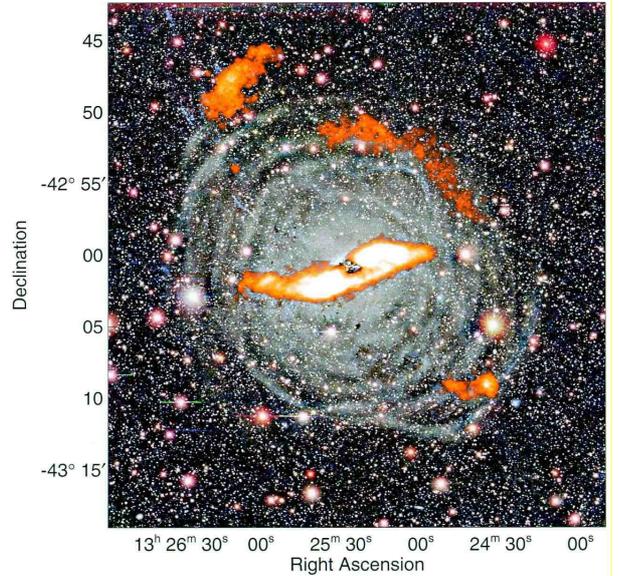,width=8cm,angle=0}}
\caption{\HI\ emission obtained from the ATCA observations (orange) superimposed to an optical BVR image obtained after unsharp masking and adaptive histogram equalisation by Peng et al. (2002).  The optical image emphasises the low-contrast features and in particular the complex set of shells and the faint dust extensions. This overlay was kindly provided by T. Oosterloo.}
\label{bello}
\end{center}
\end{figure}

\begin{figure}
\begin{center}
\centerline{\psfig{figure=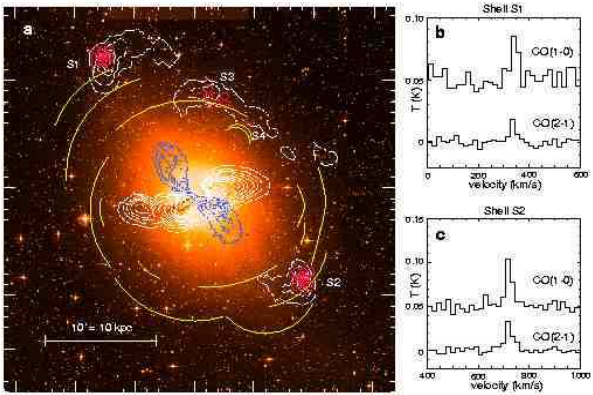,width=8cm,angle=0}}
\caption{Figure taken from Charmandaris et al. (2000) showing (a) the  locations of the various gas components and (b-c) CO(1-0) and CO(2-1) spectra toward the northern shell (upper plot) and toward the southern shell (lower plot). In (a) DSS optical image of \cena\ with the (white) contours of the \HI\ gas (Schiminovich et al. 1994). The positions observed in CO by  Charmandaris et al. (2000) are marked by circles while the location of the outer stellar shells are underlined by the yellow solid lines. The inner radio lobe is represented by the blue contours (from Clarke et al. 1992).}
\end{center}
\end{figure}
\label{COlarge}

All this is even more interesting considering that groups of young stars  have been found in the same region. The  estimated ages of these stars are about 15 Myr (see Mould et al.\ 2000 and  Rejkuba et al.\ 2002).  Recent {\it GALEX} data (Neff et al.\ 2003 and these Proceedings) also indicate the presence of ultraviolet emission that appears to be related not only to the optical filaments, but also to the chain of young stars.   Graham (1998) suggested for the first time the possibility - discussed later by many other authors - that the interaction between the radio jet and the ISM could also trigger star formation. The signature of an on-going interaction seen in the kinematics of the \HI\ appears to further support this idea. Oosterloo \& Morganti (2005) have suggested that gas stripped from the \HI\ cloud would give rise to both the large filament of ionised gas and the star formation regions that are found downstream from the location of the interaction.  Given the amount of \HI\ with anomalous kinematics and the current star formation rate, the efficiency of jet-induced star formation is at most of the order of a percent.  

As a final remark  on the ionised gas found in the filaments, it is worth mentioning that regions of ionised gas at  large distances from the nucleus are observed in many other radio sources (Baum et al. 1988). 
However, some cases appears particularly similar to \cena. Examples of these objects are PKS~2152-69 (Fosbury et al. 1998) and IC~2497 (so-called "Hanny's voorwerp", J\`ozsa et al. 2009). 

As mentioned above, \HI\ has been detected extending up to $\sim15$ kpc from the centre (Schiminovich et al. 1994, Oosterloo \& Morganti 2005) distributed in a structure that appears consistent with a partially rotating ring with the same sense of rotation  as the main body of the galaxy. 
It is intriguing that some of the \HI\ is roughly associated with optical shells (on the northern side of the galaxy, Schiminovich et al. 1994), see also Fig. \ref{bello}. The presence of \HI\ in the shells is not expected: the dynamics of the gas and stellar component  is expected to decouple during a merger event and, due to dissipation, the gas rapidly concentrates in the nucleus and, unlike stars, does not form  shell. 
A possible explanation is that the structure is the result of the  interaction between the \HI\ and the X-ray halo. Indeed \cena\ has an  extended halo with as much as 50\% of the X-ray coming from a hot corona (Forman, Jones \& Tucker 1985). The corona extends out to the \HI\ associated with the shells (see the overlay in  Fig. 5 in Schiminovich et al. 1994), suggesting the possibility of an interaction between the  hot  and the cold gas.  

It is interesting to note that  molecular gas has also been found at the location of the shells (Charmandaris et al. 2000, see Fig. \ref{COlarge}). Surprisingly, the location of the molecular gas is offset compared to  the location where the young stars are found and no young stars seem to be present in coincidence with the CO. The ratio M(\HI)/MH$_2$  is, in these regions,  close to unity and similar to what found in the central regions.  

Charmandaris et al. (2000) suggest that the molecular gas could be originating from the merger  and,  while the diffuse atomic gas at  these distances is expected to lose its energy and fall toward the centre, much denser ($10^3$ cm$^{-3}$) molecular clouds may be expected to survive at these locations. On the other hand,  the presence of CO could help explain the atomic neutral hydrogen at the location of the shells: the \HI\ would represent the diffuse envelope of  dense molecular clouds, and this combination would prevent the \HI\ to be quickly driven toward the centre during the  formation of the shell structures.

\begin{figure*}
\centerline{\psfig{figure=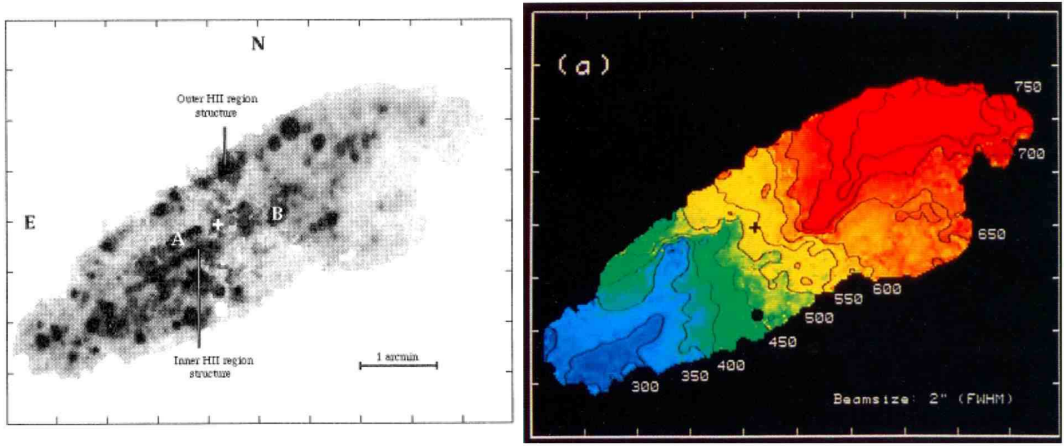,width=14cm,angle=0}}
\caption{Total intensity and velocity field of the  H$\alpha$  emission line obtained from the TAURUS imaging Fabry-Perot system at the AAT (Bland et al. 1987, Nicholson et al 1992). The images were kindly provided by Joss Bland-Hawthorn.}\label{Halphadust}
\end{figure*}

The partial ring structure of the \HI\ clouds  (Schiminovich et al. 1994) further supports the external origin of the gas in \cena.  
Furthermore, recent deep ATCA \HI\ observations  have revealed two more small clouds  of \HI\ located at about 5 and 10 kpc from the centre in the NE region (see below and Struve et al. 2010a). One cloud is observed in emission and one is seen as \HI\ absorption against the radio continuum. This result suggests that more \HI\ could be present. 
The existence of additional, low mass ($<10^6$~M$_{sun}$) \HI\ clouds is not surprising. The velocities of the two newly discovered \HI\ clouds are in agreement with the spatial velocity distribution (gradient) of the \HI\ ``outer ring'' structure and their very narrow \HI\ profile,  suggests that these clouds are more likely connected with the large scale \HI\ structure and more likely the left over of the interaction/merger discussed above. 

It is interesting to note that the \HI\ in the outer ring combined with the \HI\ in the dust-lane disk (see below)  suggests a flat rotation curve  (250 \kms\ at 15 kpc from the centre). Using this, a value of  $M/L \sim 6$ is derived (Schiminovich et al. 1994). This value is consistent with what found for other early-type galaxies from  \HI\ observations (see e.g. Morganti et al. 1999b). However, a  comparison with values derived from the X-ray halo (Forman, Jones \& Tucker 1985)  shows that the $M/L$  derived from the X-ray is at least a factor 4 higher. This discrepancy has not been explained, except by invoking the effect of discrete sources in the X-ray luminosity. 

Very interesting is also the distribution of the hot gas.  The overall X-ray structure of \cena\ was first studied by Feigelson et al. (1981) using  {\it Einstein IPC} data. These data already revealed how complex also the X-ray emission is in this object. Feigelson et al. (1981) identify emission from different structures in addition to the smooth diffuse halo of hot gas. These structures include the radio jet, confirming the early detection of Schreier et al. (1979).
The X-ray structures have been further studied with exquisite details thanks to deep {\it Chandra} and {\it XMM-Newton} observations (see Kraft et al. 2009 and ref. therein).  
These observations have shown that the X-ray emission from the inner jet and radio lobes display significant differences between the NE lobe and the SW lobe. The dominant X-ray emission mechanism from the inner jet is synchrotron radiation emanating from $\sim 30$ knots embedded in diffuse emission. There are signiÞcant spatial offsets between the X-ray and radio peaks of the inner jet (Kraft et al. 2002), and proper motions of some radio knots show velocities of $\sim 0.5 c$ (Hardcastle et al. 2003). 
A shell of shock-heated gas and X-ray synchrotron emitting ultra-relativistic electrons surrounds the SW lobe (Kraft et al. 2003, 2007; Croston et al. 2009). The temperature and density of the gas in the shell is several times that of the ambient medium. Thus, the inflation of the SW lobe is driving a strong shock into the ISM. Surprisingly, there is no corresponding shell of shock-heated gas around the NE lobe.

Feigelson et al. (1981) also reported the discovery of an X-ray filament along the south-east edge of the North Middle Lobe. This feature was also detected in EXOSAT (Morini et al. 1989), ASCA (Morganti et al. 1999), and in unpublished archival ROSAT PSPC observation. Feigelson et al. (1981) argued that this feature was probably from hot gas, and rejected both synchrotron emission from a population of ultra-relativistic electrons and inverse-Compton scattering of CMB photons as viable possibilities.
A deep image of this region has been obtained by Kraft et al. (2009) using XMM-Newton. They find that the X-ray filament consists of five spatially resolved X-ray knots embedded in a continuous diffuse bridge. 
Based on energetic arguments, they concluded that it is implausible that these knots have been ionised by the beamed flux from the active galactic nucleus of \cena\  or that they have been shock-heated by supersonic inflation of the North Middle Lobe. Instead, the most viable scenario for the origin of the X-ray knots appears to be that they are the result of cold gas shock heated by a direct interaction with the jet. 

All this clearly illustrate how complex the interaction between the radio jet and the ISM can be, even well outside the nuclear regions.

\begin{figure*}
\centerline{\psfig{figure=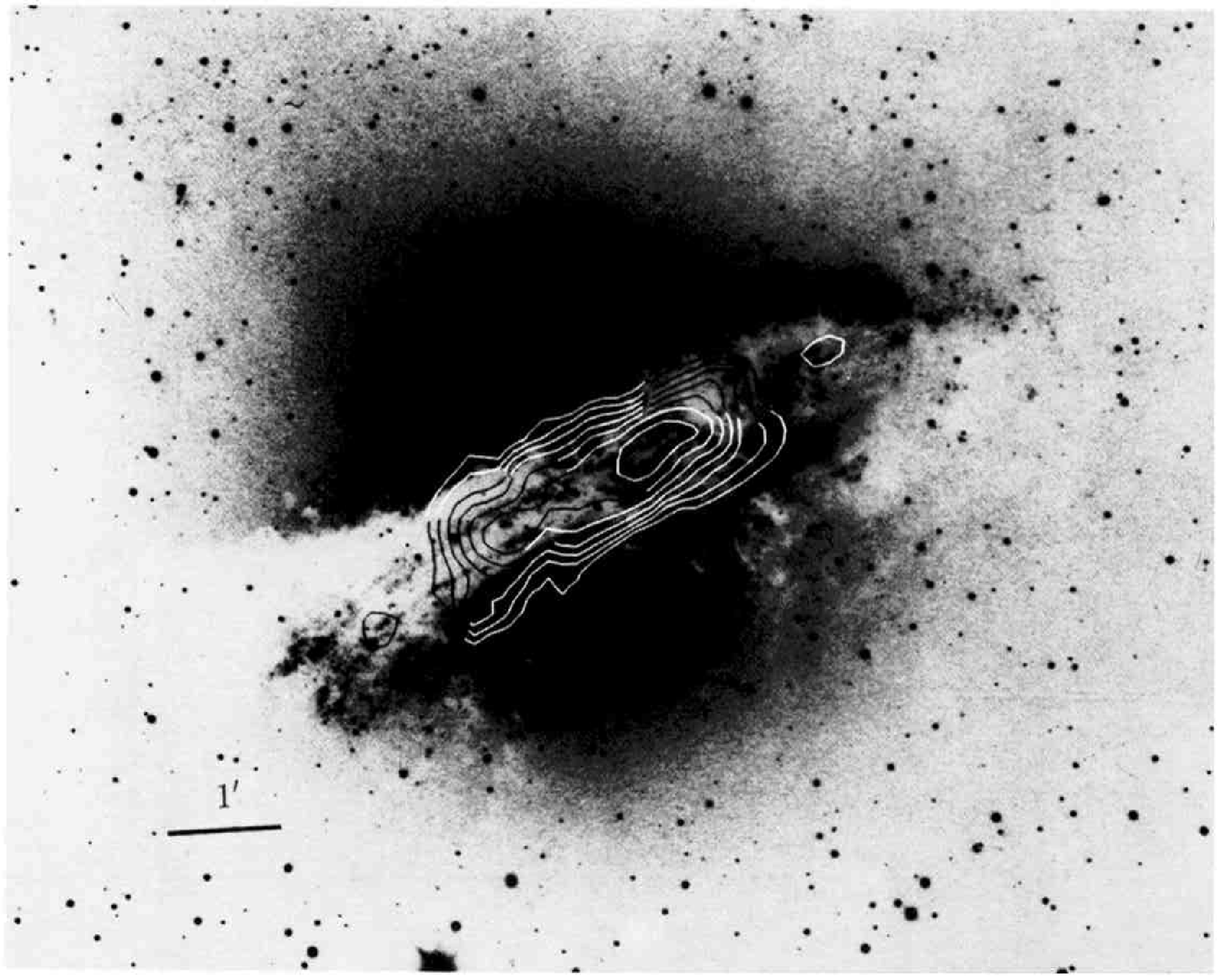,width=8cm,angle=-90}
\psfig{figure=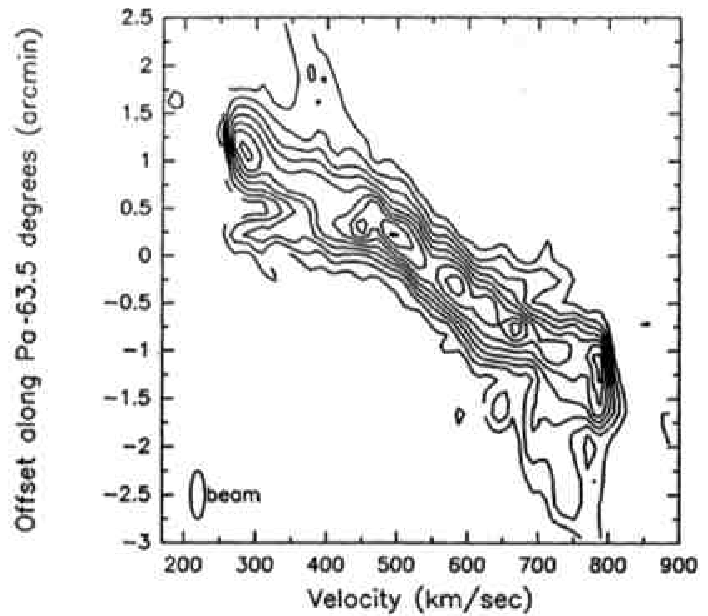,width=8cm,angle=0}}
\caption{({\sl Left}) Contour map of the integrated $^{12}$CO(1-0) emission in \cena\ superimposed to an optical image (taken from Eckart et al. 1990).  ({\sl Right}) Position-velocity map of the $^{12}$CO(2-1) from Caltech Submillimeter Observatory by Quillen et al. (1992). The plot represents a cut taken along the dust lane.  The velocity is given with respect to the systemic velocity. }\label{COdust}
\end{figure*}

\section{The gas along the dust-lane}
 
Most of the gas (ionised, atomic neutral and molecular) observed in \cena\ is actually located in a disk that follows the dust-lane.  In this region,  the distribution of the dust has been observed  by the {\it Spitzer} Infrared Array Camera  (Leeuw et al. 2002, Quillen et al. 2006), by {\it ISO} with ISOCAM and by SCUBA (Mirabel et al. 1999).

The emission from the ionised gas (and in particular H$\alpha$ emission line) has been studied using the integral field unit TAURUS on the Anglo Australian Observatory (AAO) by Bland et al. (1987) and Nicholson et al. (1992). Fig. \ref{Halphadust}  illustrates the distribution and kinematics of the ionised gas.  Nicholson et al. (1992)  presented geometrical, tilted-ring models that successfully reproduce not only the appearance of the dust-lane system seen in broad-band optical images but also the main kinematics and morphological features seen in the H$\alpha$ emission.
The structure of the warped disk  has been confirmed also by the study of the molecular gas.
Eckart et al. (1990) observed \cena\  with SEST in the  $^{12}$CO (1-0, 2-1) and $^{13}$CO lines with a  resolution of about  45$^{\prime\prime}$  and 22$^{\prime\prime}$  respectively.
Quillen et al. (1992) observed  the $^{12}$CO (2-1) line using the Caltech Submillimeter Observatory (CSO)  in order to obtain a better velocity resolution. The mass estimated for the molecular gas is $4 \times 10^{8}$ M$_\odot$. Fig. \ref{COdust} shows the distribution of the CO and its kinematics along the dust lane. 
The models of Nicholson et al. (1992) and Quillen et al. (1992) describe the overall structure observed (from the dust, ionised and molecular gas) as a transient, warped and thin disk composed of a series of inclined connected rings undergoing circular motion. The disk is rapidly rotating with a rotation gradient of about 150 \kms\ kpc$^{-1}$ and highly inclined.
Expanding on the work of Nicholson et al. (1992) and Quillen et al. (1992), Quillen et al. (1993) has suggested a timescale of about 200 Myr since the core of an infalling spiral galaxy reached and merged with \cena. Because of this encounter, an initially flat disk, misaligned with the galaxy principal symmetry axis, becomes increasingly corrugated as function of time.  
An alternative model - a polar ring model - to explain the warped structure has been instead proposed by Sparke (1996).  This dynamical model shows that the geometry of the disk can be explained as that of a near-polar structure precessing about the symmetry axis of an approximately oblate galaxy potential which is nearly round at the centre (like the stellar body).  The outer \HI\ structure described in the previous session (Schiminovich et al. 1994) can be explained as the continuation of the same precessing disk. However, the time-scales required by this model to account for the twist in the warp are longer. A summary of  the time scales derived from various diagnostics  will be given in Sec. 6.

Another important component of the gas that is present along the dust disk of \cena\ is the neutral hydrogen. The distribution and kinematics of the \HI\ along the dust-lane was first observed using the VLA  by  van Gorkom et al. (1990). \HI\ is detected mostly in emission  but also in absorption against the southern radio lobe,  the core and the beginning of the northern jet.  The morphology and kinematics of the \HI\ confirmed the warped structure although given the low spatial and kinematic resolution of this data no detailed model was performed.
The total mass of the \HI\ disk  is $3.9\times 10^8$ M$_\odot$.

More recent observations were obtained with the ATCA, achieving relatively high spatial  ($\sim 6$$^{\prime\prime}$) and 
velocity resolution. This has allowed more detailed imaging of the kinematics of the gas (and better separate the emission and absorption components).  See Struve et al. (2010a) and Struve et al. (2010b) for more details.
The quality of the data has allowed a detailed modelling using a tilted ring model of the disk. The  kinematics of the \HI\  can be described - down to the nuclear scales - as a regularly rotating, highly warped structure. Large-scale radial motions do exist, but they are only present at larger radii. This unsettled gas is mainly part of a tail/arm-like structure, possibly wrapped around the radio lobe as it appears  both in emission and absorption in the region below the dust lane. The relatively regular velocity of the gas in this structure suggests that it is in the process of settling down into the main disk. 
The parameters derived from the  tilted ring modelling of the \HI\ are similar but not identical to the one obtained by Quillen et al. (2006) while they agree very well with the results from the kiloparsec-scale stellar ring discovered by Kainulainen et al. (2009).  Furthermore, these parameters give  a good description of the Spitzer far-IR  data from Quillen et al. (2006).
There are no indications of {\sl large-scale}  anomalies in the kinematics all the way down to the central beam ($\sim 100$ pc). The presence of a large-scale bar structure  is also excluded by these data. 


It is interesting to note that a large-scale bar structure was suggested from the dust structure imaged with ISO (Mirabel et al. 1999). This structure is clearly ruled out by the more recent Spitzer far-IR data  (see Quillen et al. 2006) and not required to explain the \HI\ data (Struve et al. 2010b). However,  the discussion about the possible presence of a bar structure (in the very inner regions) and the connection to fuelling of the AGN is still ongoing  (see Sec. 4). The possible presence of such a structure has relevance for the fuelling of the AGN.

Finally, in addition to the warm and cold gas, arcs of hot gas have been detected in soft X-rays with Chandra (Karovska  et al. 2002) extending $\sim 8$ kpc in a direction perpendicular to the jet. The thermal gas would have a density between a few $\times 10^{-3}$ and $10^{-2}$ cm$^{-3}$.  A few hypotheses about the origin of this structure have been put forward. This emission may originate  from interaction (and shocks?) between infalling gas/tail from merger and the material in the host elliptical galaxy.  Alternatively,  the origin could be related to the nucleus. The emission would be resulting from an interaction between a powerful outflow (or  wind) from the nucleus and the system of stellar shells or the ISM. This giant eruption/outburst would have taken place ~10$^7$ yrs  ago (by deriving a mean velocity of the ejecta of $\sim 600$ \kms, Karovska  et al. 2002). 

\section{The sub-kpc region}

Various studies have pointed out the presence of a gap in the distribution of gas  and dust between 10$^{\prime\prime}$ and  45$^{\prime\prime}$ (from ~0.2 and  0.8 kpc). This has been observed in H$\alpha$, CO and, recently, in \HI\  (Nicholson 1992, Quillen et al. 2006 and  Struve et al. 2010b, respectively). 
As also discussed in Struve et al. (2010a) ring-like structures are, in fact, seen in a number of early-type galaxies (see Donovan et al. 2009 and ref. therein). Various possibilities for the  origin of  this gap have been considered. In particular, the well-studied case of IC~2006 (Schweizer, van Gorkom \& Seitzer 1989) shows that gas accretion can also form such structures. In addition to this, effects of inclination or effects related to nuclear energetic processes should also been considered.  
 
\begin{figure}
\centerline{\psfig{figure=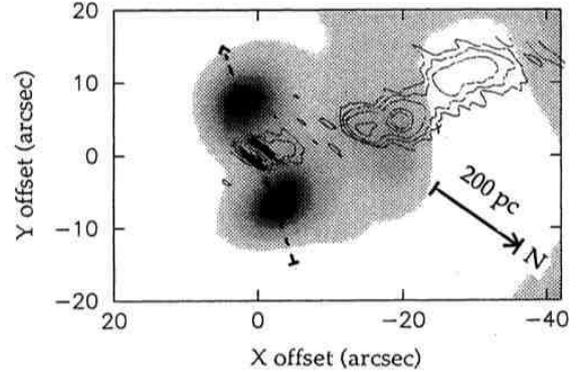,width=8cm,angle=0}}
\caption{Deconvolved image (grey scale) of the CO (2-1) obtained by using only the highly red- and blue-shifted gas (from Rydbeck et al. 1993). The contours represent the 1.5~GHz VLA image of the radio jet from Burns et al. (1983)}\label{rydbeck}
\end{figure}

On the other hand, gas is present in the nuclear regions of \cena, i.e.  inside the inner ~200 pc. The presence of a circumnuclear ring of molecular gas  has been inferred for the first time from CO data, thanks to the detection of broad, asymmetric  Òhigh velocity wingsÓ that were interpreted as due to rapidly rotating molecular gas close to the nucleus (Israel et al. 1990). 
A CO intensity image obtained using only these high velocity wings shows indeed an edge-on ring radius of ~70 -120 pc   perpendicular to the radio jet (see Fig. \ref{rydbeck} and Rydberk et al. 1993). This disk is barely resolved in the $^{12}$CO(3-2) data of Liszt (2001) obtained with the {\it JCMT} at about 14$^{\prime\prime}$ resolution.  

\begin{figure}
\centerline{\psfig{figure=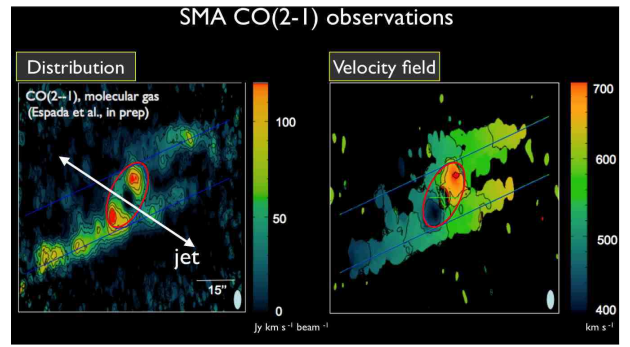,width=8cm,angle=0}}
\caption{12CO(2-1) total intensity (left) and velocity field (right) taken from Espada et al. (2009), Submillimeter Array observations. The colour scale ranges from 400 \kms\ up to 700 \kms.}\label{espada}
\end{figure}

\begin{figure*}
\centerline{\psfig{figure=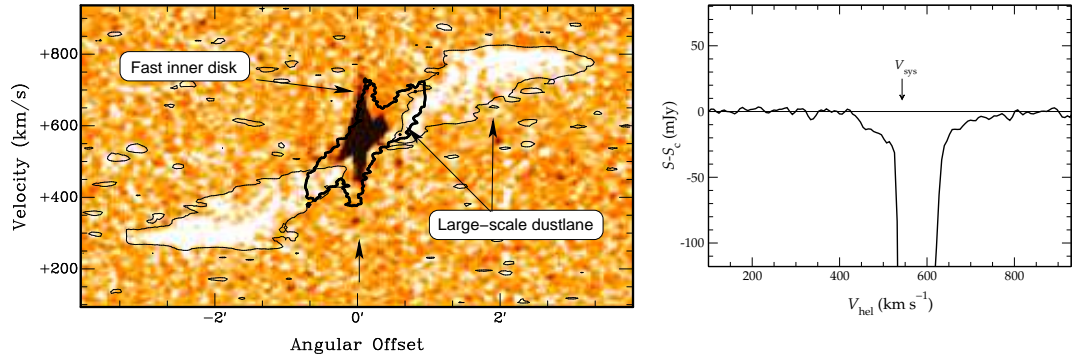,width=14cm,angle=0}
}
\caption{{\sl left:} Position-velocity plot of the \HI\ (grey-scale and thin contours) and,
superimposed, the CO emission (thick contours; from Liszt 2001), taken along position angle 139$^\circ$.  The gray-scale represents the high-resolution \HI\ data (beam 8$^{\prime\prime}$) while the thin contour is from the same dataset smoothed to 15$^{\prime\prime}$.  Note that the CO observations do not extend beyond a radius of about 1 arcmin.
{\sl right:} The absorption profile against the central radio core (indicated by the arrow in the left panel)  showing the blue- and red-shifted wings of the absorption profile.}\label{HIabs}
\end{figure*}

This inner molecular disk  has been recently observed, with arcsec resolution, in $^{12}$CO(2-1)  by  Espada et al. (2009) using the Submillimeter Array (SMA). These observations spatially resolve the circumnuclear molecular gas in the inner few hundred pc (400 pc $\times $  200 pc), revealing that this disk is elongated along a position angle $\sim 155^\circ$ and, therefore, perpendicular to the radio/X-ray jet. The southeast  and northwest  components  of the circumnuclear gas are connected to molecular gas found at larger radii. See Fig. \ref{espada}.

Espada et al. (2009)  have extened the available models to the inner radii ($r < 200$ pc)  in order to reproduce the 
central parallelogram-shaped structure observed in $^{12}$CO(2-1), see Fig. \ref{espada}. Adopting the warped disk model, they confirmed a gap in emission between 200 and 800 pc radii,  as mentioned above.  However, they also found spatial and kinematical asymmetries in both the circumnuclear and outer gas, suggesting non-complanar and/or non-circular motions. They propose a model combining a weak bar in a warped disk. However, the velocity field derived by their proposed model does not show the large gradient seen in the data in the intersection between the circumnuclear gas and the gas at larger radii, see  Fig. \ref{espada} and Fig.~11 of their paper. More work will need to be done to establish whether a weak bar structure is present in the very centre of \cena.

The nuclear regions have also been explored with near-IR spectroscopy.  Marconi et al. (2001) have presented long slit near-IR spectroscopy of the ionised and molecular gas ([FeII], Pa$\alpha$, Br$\gamma$, H$_2$)  in the inner $\sim 200$ pc.  They identify three components. A large scale disk, a "ring like" system with inner radius 6$^{\prime\prime}$ and detected only in H$_2$ and, finally, the likely counterpart of the 100 pc disk seen in CO. However, using these data they could not distinguish between the presence of  a bar structure or a warped disk for this inner region of the gas.  No optical emission from star formation associated with the radio jet was detected (Marconi et al. 2001 ). 
Expanding on these results, higher resolution near-IR  observations using AO have allowed a better insight in the very inner regions and will be discussed  in the next session.

Toward the central continuum regions of \cena, many atomic and molecular species have been detected  thanks to observations of absorption lines. In particular, \cena\ was the first object in which \HI\ absorption was detected (Roberts 1970) using single dish observations.  Higher resolution observations first suggested  that the \HI\ absorption against the nuclear region was  mostly redshifted (van der Hulst et al. 1983, van Gorkom et al. 1989). The conclusion from these observations was that the \HI\ gas was mainly infalling and likely feeding AGN. 
However, the deeper ATCA data described  above have shown that the situation is more complicated than this.  Absorption has been detected against the radio core also at velocities blueshifted with respect to the systemic velocity (Morganti et al. 2008).
The nuclear \HI\ absorption appears asymmetric with respect to the systemic velocity (taken as 542 \kms\ from van Gorkom et al. 1990), with  velocities ranging from $\sim$$400$ \kms\ (about $-140$ \kms\ blueshifted compared to  systemic) up to $\sim$$800$ \kms\ (i.e., about +260 \kms\ redshifted relative to  systemic), see Fig. \ref{HIabs}. This absorption profile is, therefore,  much broader 
than reported before. With these new results, the kinematics of the  \HI\ in the inner regions of \cena\  appears very similar to that observed in emission for the molecular circumnuclear disk. In particular, Fig.\ref{HIabs} shows how the \HI\ absorption compares with the CO disk observed by  Liszt (2001). It is important to note that the modelling of the large-scale \HI\ has allow to rule out that the absorption is due to gas at large distances projected against the centre (Struve et al. 2010b).
Thus, these results suggest that the central \HI\ absorption does not provide  direct evidence of gas infall into the AGN, but instead is (at least partly) due to a cold, circumnuclear disk (Morganti et al. 2008). This also rules out the presence of nuclear fast outflows in \cena, unlike what has been detected in other radio galaxies. 
In \cena, the effects of the interaction between the radio plasma and   ISM seem to happen on the larger scale (i.e. inner radio lobes, Croston et al. 2009 or Northern Middle Lobe).
However, it is not yet clear what is the illuminating background continuum source at this frequency against which the \HI\ absorption is detected.  As shown in  Fig. \ref{neumayer} (left), the radio core is already not detected at 2.3 GHz.    
Indeed, as shown by Jones et al. (1996) and  Tingay \& Murphy (2001), between 2 and 5~GHz the innermost part of the radio continuum source is affected  by free-free absorption that  might be caused by circumnuclear ionised gas. The absorption of hard X-rays, indicating a column density of $10^{23}$ atoms cm$^{-2}$  of absorbing gas in front of the central source (Evans et al. 2004), also suggests the presence of circumnuclear material around the super-massive black hole, as expected from the AGN unification models. Thus, the \HI\ absorption is occurring either against the counter jet or against a more extended region, e.g. due to
thermally emitting ionised gas evaporated from the inner edge of a torus or disk.  Deep, high resolution \HI\ observations will be needed to shed light on this.

Many different molecular lines have been detected against the nucleus in absorption: H$_2$CO, OH, NH$_3$, C$_3$H$_2$ (Gardner \& Whiteoak 1976, Seaquist \& Bell 1986, 1990, Israel et al. 1990).  For a more complete summary of the detected species see Israel (1998) and Muller \& Dinh-V-Trung (2009). As in the case of the \HI, it is still a matter of debate whether these tracers are located in the circumnuclear regions. 

\begin{figure}
\centerline{\psfig{figure=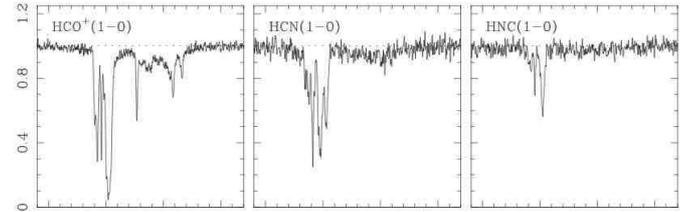,width=9cm,angle=0}}
\caption{Spectra of the observed transitions detected by Wiklind \& Combes (1997).  The spectra have been normalised to a continuum level of unity, see Wiklind \& Combes (1997) for details.}\label{absmolecules}
\end{figure}

\begin{figure*}
\centerline{\psfig{figure=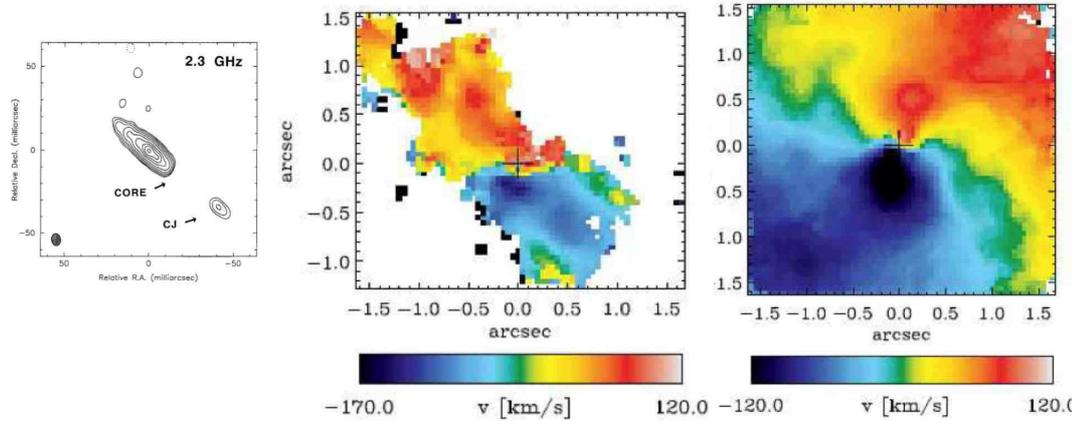,width=14cm,angle=0}}
\caption{Panel showing (left) the 2.3~GHz VLBI jet with the location of the (free-free absorbed) core  (Jones et al. 1996); (middle) the velocity map of the [SiVI] line emission with teh velocity scale given relative to the systemic velocity of the galaxy  (assumed 532 \kms) from Neumayer et al. (2007); (right) same for as middle pannel but for H$_2$ (Neumayer et al. 2007).}\label{neumayer}
\end{figure*}

HCO$^+$ and CO weak absorption features were detected by Israel et al. (1991) at redshifted velocities compared to the systemic velocity and  explained as gas falling into the nucleus.  More observations of  the HCO$^+$ and other lines were  published by Wiklind \& Combes (1997) and Fig. \ref{absmolecules} shows  examples of their molecular absorption profiles. These profiles can be decomposed in a series of deep, very narrow components (with a width of 1-2 \kms) located close to the systemic velocity and a broad component.  This is, to first order, similar to the \HI\  profile although no one-to-one correspondence of the narrow components has been found  and the broad component  is mostly redshifted  in the molecular gas although a hint of very low optical depth blueshifted HCO$^+$ absorption wing is also seen (Israel et al. 1992 and Wiklind \& Combes 1997).  The broad component has been explained by these authors as perhaps originated by a circumnuclear disk. 
An alternative scenario has been proposed by  Eckart et al. (1999): the general velocity stucture of the absorption could be explained kinematically with a tilted-ring model and high latitude clouds, not necessarely requiring the presence of molecular gas close to the active nucleus. However, the thickness would not be enough to explain e.g. the range of velocities detected in \HI. 

The results from VLBI observations of the OH transitions (van Langevelde, Pihlstr\"om \& Beasley 2004), as well as the excitation 
modelling, appear to be also consistent with the absorption occurring on 200 Ð 2000 pc 
from the center of \cena. On the other hand, the  formaldehyde absorption detected on VLBI scale seems to have a distinctly  different distribution, possibly restricted to small, high density  clouds in the circumnuclear disk (van Langevelde et al. 2004).

In general, it is clear that the physical conditions of the absorbing molecular gas are still not completely understood.
For example, recent observations of HCN$^+$ and HCN by Muller \& Dinh-V-Trung (2009) found no evidence for molecular gas density higher than a few $\times 10^4$ cm$^{-3}$ (unlike previous claims), thus suggesting that either the line-of-sight to the radio continuum does not intercept the circumnuclear disk  or the density of the absorbing gas in the circumnuclear disk is lower than $10^4$ cm$^{-3}$. 

\section{The very inner region ($< 30$ pc)}

Ionised and molecular gas in the inner $\sim 30$ pc have been studied using integral field spectroscopy by Krajnovic et al. (2007) and Neumayer et al. 
(2007).  The latter made use of SINFONI AO-assisted at the VLT to achieve the highest possible spatial resolution.  Neumayer et al. 
(2007) found that for higher excitation lines [SiVI] and [FeII] the velocity pattern is increasingly dominated by a non-rotating component, elongated along radio jet. Interestingly, these non-rotational motions were detected in the direction {\sl along} the radio jet, with redshifted velocities (compared to the systemic) seen on the main-jet side  and blushifted on the counter-jet side. These motions (stronger in [SiVI]) can be explained as backflow of gas accelerated by the plasma jet. 

On the other hand, the velocity of the  H$_2$ line is dominated by rotation, although with a quite high dispersion ($\sigma \sim 400$ \kms\ for the H$_2$ line and higher for the other lines). The reason for this high value of the velocity dispersion is not clear but  partially resolved rotation or  local turbulence have been considered. 
All in all, the H$_2$ data in Neumayer et al. (2007) appear to be fully described by a tilted-ring model and gas moving in circular orbits. On such a small scales, the origin of the structure of the nuclear disk could be due to self-induced warping of the accretion disk.

Although the overall shape of the H$_2$ velocity field appears to be point-symmetric, Neumayer et al. (2007) point out some asymmetries of the peak velocities in the field. These asymmetries may be related to the fuelling process of the nuclear disk suggested by the morphology and kinematics of the high ionisation lines. Interestingly, the range of velocities covered by the \HI\ absorption against the nuclear regions  (Morganti et al. 2008) is similar to the range covered by the near-IR lines on the blueshifted side, while they are larger on the redshifted side (extending to $\sim 800$ \kms\ while the H$_2$ and [SiVI] cover only up to $\sim 650$ \kms). 

One of the results from these IFU studies is also the determination of the black-hole mass in \cena.  From the Neumayer et al. (2007) tilted-ring model of the gas, a BH mass $4.5 \times 10^7$ M$\odot$  has been derived (for an inclination $i =34^\circ$). This is lower than found in previous studies. The difference (and improvement) is attributed to the fact that lines with more  regular rotation (i.e. less affected by the radio jet) have been used. Interestingly, a similar BH mass has been obtained from the stellar kinematics (Cappellari et al. 2009).  The new estimated value of the BH nicely fits in the M$_{\rm BH} - \sigma$ correlation (see Fig. 17 in Neumayer et al. 2007).  

Finally, it is worth mentioning that, using X-ray data, Evans et al. (2004) estimated the accretion rate  (Bondi accretion) of gravitationally captured hot gas into the central BH to be   $\sim 6.4 \times 10^{-4}$  M yr$^{-1}$ and  a corresponding efficiency of  $\sim$ 0.2\%. These values (although derived  using old values of the BH mass and, therefore, requiring an update) suggest that \cena\ has  intermediate characteristics. The value of the accretion rate is  higher than in galaxies with radiatively inefficient fuelling of the AGN while smaller than for a canonical efficient accretion system. 
Nevertheless, as mention above, the high column density  as well as the presence of Fe Ka line derived from the X-ray observations   (Evans et al. 2004) support the idea  that large quantities of cold gas located in a molecular torus at about 1pc from the BH are present in \cena.

\section{Time scales and concluding remarks }
 
Gas structures have been observed in \cena\ ranging from many tens of kpc to sub-pc scales.
The external origin of the gas via accretion of a small, gas-rich galaxy is supported by some of the observed characteristics,                                                                                                                                                                                                                                                        like the  misalignment of the rotation axis between stars and gas and the presence of  shells.  Although this merger may be the origin of the warped structure of the disk,  the gas in the disk appears now  well settled in a regular structure (warped thin disk and on circular orbits) down to the nuclear regions.  

Given the variety of diagnostics that can be observed in \cena, it is interesting to put together some of the time scales involved in producing the observed structures:
\begin{itemize}
\item time scale  to create shells from merger: between 2 and $6 \times 10^8$ yrs  (Quillen et al. 2006);
\item time scale (to form a warp) from the precessing model:  $\sim 7\times 10^8$ yr (Sparke 
1996); 
\item time scale to create the warp $\sim  2 \times 10^8$ yrs (e.g. from CO and Spitzer data, Quillen et al. 1992, 2006);
\item time scale obtained from the structure and kinematics (and the modelling) of the \HI\ disk:  $1.6 - 3.2 \cdot 10^8$~yr since the merging event   (van Gorkom et al. 1990, Struve et al. 2010a); 
\item ages of stars  associated with the young blue tidal stream: 300 Myr (Peng et al. 2002);                                                     
\item ages of young stars and star clusters  in the centre:  between $10^6$ to few times $10^7$ yrs (Dufour et al. 1979; Moellenhoff 1981, Minniti et al.  2004);
\item much longer ages - of the order of a few Gyr - are derived from Planetary Nebulae  and Globular Clusters  (Peng et al. 2004a,b) 
\item age of the large radio structure (e.g. NML):  estimated to be  a few $\times  10^7$ yrs  from the still ongoing particle injection, Hardcastle et al. 
(2008)  and up to  $10^8$ yrs if the north middle radio lobe is described as a rising buoyant bubble structure(Saxton et al. 2001) 
\item age of the inner radio lobe:   few $\times  10^6$ yrs from X-ray emission (Croston et al. 2009)
\end{itemize}

Thus, the time scales derived for the formation of the warped disk structure are much longer than the time scale of the inner radio lobes while more similar to the time scales derived for the middle radio lobe (see Sec 6). This rises the question of whether  the merger/accretion  and AGN activity are actually at all related.  

The fuelling mechanism of the nuclear activity is also not completely clear.  However, the very high resolution of near-IR observations points to evidence for some infalling gas that may do the work.  There seems to be no obvious evidence for outflows in the nuclear regions while there is evidence on the larger (kpc) scale of interaction between the radio jet and the ISM.  These lobes are only 6~kpc  in radius, so this may represent what commonly happens in young or restarted radio sources when they reach the kpc scales. This would, therefore, confirm how the early stages of radio-galaxy evolution potentially represent an important galaxy feedback process.

\end{document}